\documentclass[11pt,titlepage,a4paper]{article}
\usepackage{bozhomac}	
\usepackage{bozhlogo}	
\usepackage{cite}	
\usepackage{tabularx}   

\title{\bfseries	\vspace*{-1.678902345in}
{\huge  The ``\emph{Lorenz} gauge'' is named in\\[1ex]
	honour of Ludwig Valentin Lorenz!
}
}

\vspace{1.7ex}

\author{
Bozhidar Z.\ Iliev
\thanks{Laboratory of Mathematical Modeling in Physics,
Institute for Nuclear Research and \mbox{Nuclear} Energy,
Bulgarian Academy of Sciences,
Boul.\ Tzarigradsko chauss\'ee~72, 1784 Sofia, Bulgaria}
\thanks{E-mail address: bozho@inrne.bas.bg}
\thanks{URL: http://theo.inrne.bas.bg/$\sim$bozho/}
}

\date{
 \vspace{2.27ex}\ShortTitle{The ``\emph{Lorenz} gauge'' is named in honour of
						L.V.\ Lorenz!} \\[0.27ex]
 \vspace{3.27ex}
\small
	\begin{tabular}{r@{$\colon\to~$}l}
 \vspace{0.27ex} Produced	& \fbox{\today}	\\[0.27ex]
	\end{tabular} \\[1.27ex]
\normalsize
\small
	\begin{tabular}{r@{$\colon~$}l}
\normalsize\sffamily\bfseries
\vspace{0.27ex} http://www.arXiv.org e-Print archive No. &
\normalsize\sffamily\bfseries
arXiv:0803.0047 [physics.hist-ph]		\\[0.27ex]
	\end{tabular} \\[-0.27ex]
\normalsize
 \vspace{4.27ex}{\Huge\BOZHO}	\\[4.27ex]
\vspace{0.27ex}\Subject{Electrodynamics, Gauge theories}	\\[2.27ex]
	\begin{tabular}{r@{\hspace{0.512em}}|@{\hspace{0.512em}}l}
\vspace{0.27ex}\MSC[2001]{01-01, 78-03\\ 78A25, 81S99, 81T13}
&
\vspace{0.27ex}\PACS[2003]{1.65.+g, 03.50.De\\ 03.70.+k, 11.15.-q}
	\end{tabular} \\[1.27ex]
\vspace{0.27ex}\KeyWords{Lorenz gauge, Lorenz condition \\
		Electromagnetic potentials, Gauge potentials \\
		Gauge conditions, Gauges,
		Electrodynamics, Gauge theories, Yang-Mills theories} \\[0.27ex]
}

\begin{document}		

\renewcommand{\thepage}{\roman{page}}

\renewcommand{\thefootnote}{\fnsymbol{footnote}} 
\maketitle				
\renewcommand{\thefootnote}{\alph{footnote}}   



\begin{abstract}

	The letter reminds the historical fact that the known ``Lorenz
gauge'' (or ``Lorenz condition/relation'') is first mentioned in a written form
by and named after Ludwig Valentin Lorenz and not by/after Hendrik Antoon
Lorentz.

\end{abstract}

\renewcommand{\thepage}{\arabic{page}}


\section {Introduction}
\label{Introduction}

	Let $\varphi$ and $\bs{A}$ be respectively the scalar and vector
potentials of the classical electromagnetic field~\cite{L&L-2}. Since they
contain some arbitrariness, they can be connected via different relations,
called gauges or gauge conditions/relations. A particular example of them being
the ``\emph{Lorenz} gauge'', \viz
    \begin{equation*}    \label{Lorenz}
\nabla\cdot\bs{A} + \frac{1}{c} \frac{\pd \varphi}{\pd t} = 0,
    \end{equation*}
where $c$ is the velocity of light in vacua,
$\nabla:=\bigl(\frac{\pd }{\pd x},\frac{\pd }{\pd y},\frac{\pd }{\pd z}\bigr)$
and $x$, $y$, $z$ and $t$ are the space and time coordinates, respectively. In
invariant form, this gauge reads
    \begin{equation*}    \label{Lorentz-inv}
\pd^\mu A_\mu = 0
    \end{equation*}
with $A_\mu$, $\mu,\nu=0,1,2,3$ being the components of the 4-vector potential
and $\pd^\mu=\eta^{\mu\nu}\frac{\pd }{\pd x^\nu}$, where $\eta^{\mu\nu}$ are
the components of the the standard metric tensor of the Minkowski spacetime and
$(x^0,x^1,x^2,x^3)=(ct,x,y,z)$.

	More generally, if $A_\mu^a$, $a=1,\dots,n\in\field[N]$, are the
4-potential of an arbitrary classical gauge field, they can be subjected to the
Lorenz relation in the form
    \begin{equation*}    \label{Lorentz-gauge}
\pd^\mu A_\mu^a = 0 .
    \end{equation*}

	The importance of the Lorenz gauge comes from its relativistic
invariance, from a simplification of many calculations in it, etc.

	A partial analysis of the Lorenz gauge in quantum electrodynamics can
be found in~\cite[ch.~II, \S~12]{Bogolyubov&Shirkov} or~\cite[ch.~9,
\S~2]{Schweber} (see also~\cite[ch.~I, \S~5]{Akhiezer&Berestetskii}).


\section{The historical truth}
\label{Sect2}

	The Lorenz condition/relation and gauge are named in honour of the
Danish theoretical physicist Ludwig Valentin Lorenz (1829--1891), who has first
published it in 1867~\cite{Lorenz/1867,Lorenz/1867a} (see also~\cite[pp.\
268-269, 291]{Whittaker-History} and~\cite{Bladel-1991,Nevels&Shin}). However
this condition was first introduced in lectures by Bernhard G.~W.~Riemann in
1861 as pointed in~\cite[p.~291]{Whittaker-History}.

	It should be noted that the \emph{Lorenz} condition/gauge is quite
often erroneously referred as the Loren\hspace{-0.0168em}\emph{t}z
condition/gauge after the name of the Dutch theoretical physicist Hendrik
Antoon Lorentz (1853--1928) as, e.g., in~\cite[p.~18]{Roman-QFT},
\cite[p.~45]{Gockeler&Schucker} and~\cite[pp.~421-422, 426, 542]{Griffiths-1998}.

	The table below represents some results of searching  over the Internet
for ``\emph{Lorenz} gauge''  and ``Lorentz gauge.'' We see that the situation
is quite sad in favour of the wrong term, but there is a slight improvement
during the last 3 years.

\renewcommand{\arraystretch}{1.32}
    \begin{table}[ht!]  \label{Lorenz-Lorentz-Table}%
	\begin{minipage}{\textwidth}
\caption{Number~%
\protect\footnote{~The number of found results depends on may factors and may
be variable even during one day.}
 of found search results for ``Loren\emph{t}z gauge'' and
			    ``Lorenz gauge''\vspace{2ex}}
    \begin{tabularx}{\textwidth}{ll@{\hspace{2.5em}}lll}
 & & & \\
\hline
Web database~%
\footnote{%
The URL of arXiv is http://arxiv.org,
of Google is http://www.google.com,
of Google Scholar is http://scholar.google.com,
of Yahoo is http://search.yahoo.com,
of AOL is http://search.aol.com,
and of Ask is http://www.ask.com.%
}
	& Date	& ``Loren\emph{t}z gauge''&\bfseries ``Lorenz gauge'' & Ratio
\\ \hline
arXiv full record (Physics) &\itshape Jul 2005 &  89 &\bfseries 17 & 5.25 \\
arXiv full record (Physics) & Feb 2008 &  85 &\bfseries 25 & 3.40 \\
Google 			    &\itshape Jul 2005 & 13700 &\bfseries  558 & 24.55 \\
Google 			    & Feb 2008 & 22600 &\bfseries 5500 &  4.11 \\
Google Scholar 		    &\itshape Jul 2005 & 2640 &\bfseries  216 & 12.22 \\
Google Scholar 		    & Feb 2008 & 5030 &\bfseries  499 & 10.08 \\[1ex]
arXiv exp. full text (Physics) & Feb 2008 & 2273 &\bfseries 345 & 6.59 \\
arXiv exp. full text (Math.)   & Feb 2008 &  133 &\bfseries  28 & 4.75 \\
Yahoo 			    & Feb 2008 & 21700 &\bfseries  4460 & 4.87 \\
AOL 			    & Feb 2008 &  6020 &\bfseries   944 & 6.38 \\
Ask 			    & Feb 2008 &  1600 &\bfseries   658 & 2.43
\\ \hline
    \end{tabularx}
	\end{minipage}
    \end{table}

 	It is remarkable that in the book~\cite[p.~58, footnote~1]{Thide} is
mentioned the Lorenz (gauge) condition, but the author continues to call it
Lorentz (gauge) condition.

	The reasons for the error in referring to the ``\emph{Lorenz} gauge'' as
``Loren\emph{t}z gauge'' are explained and analyzed, for instance,
in~\cite[pp.~670--671]{Jackson&Okun}.


\section {On the geometry of the gauge conditions}
\label{Sect3}

	It is known that the gauge potentials of a gauge field, in particular of
the electromagnetic one, are coefficients of a linear connection on a
vector bundle from geometrical point of
view~\cite{Konopleva&Popov,Gockeler&Schucker}. If a gauge field is given, its
potentials are fixed in any frame of reference. For that reason, the imposition
of some relations between them, in particular of the Lorenz gauge, leads to
restrictions on the reference frames one can invoke. To be more precise, the
class of frames in the total bundle space that can be used is narrowed so that
the corresponding gauge relations to be satisfied.


\section {Appeal instead of a conclusion}
\label{Conclusion}

	The Lorenz gauge is in current usage and seems to be in use in future.
For that reason, let us recognize the contribution of Ludwig Valentin Lorenz
in this field of physics and restore the historical truth by terming this
gauge/relation after him!


\section*{Acknowledgments}

	This work was partially supported by the National Science Fund of
Bulgaria under Grant No.~F~1515/2005.


\addcontentsline{toc}{section}{References}
\bibliography{bozhopub,bozhoref}
\bibliographystyle{unsrt}
\addcontentsline{toc}{subsubsection}{This article ends at page}

\end{document}